\documentclass[longauth,printer]{aa}  
\usepackage{graphicx}

\newcommand{\kms}{km~s$^{-1}$}

\newcommand{\doceCO}{\mbox{$^{12}$CO}}
\newcommand{\doce}{\mbox{$^{12}$CO}}
\newcommand{\trece}{\mbox{$^{13}$CO}}
\newcommand{\treceCO}{\mbox{$^{13}$CO}}

\newcommand{\jdn}{\mbox{$J$=10$-$9}}
\newcommand{\jdsq}{\mbox{$J$=16$-$15}}
\newcommand{\jsc}{\mbox{$J$=6$-$5}}
\newcommand{\jct}{\mbox{$J$=4$-$3}}
\newcommand{\jtd}{\mbox{$J$=3$-$2}}

\newcommand{\jdu}{\mbox{$J$=2$-$1}}
\newcommand{\juc}{\mbox{$J$=1$-$0}}
\newcommand{\gsim}{\raisebox{-.4ex}{$\stackrel{>}{\scriptstyle \sim}$}}
\newcommand{\lsim}{\raisebox{-.4ex}{$\stackrel{<}{\scriptstyle \sim}$}}

\newcommand{\mloss}{\mbox{$\dot{M}$}}
\newcommand{\my}{\mbox{$M_{\odot}$~yr$^{-1}$}}
\newcommand{\ls}{\mbox{$L_{\odot}$}}
\newcommand{\ms}{\mbox{$M_{\odot}$}}

\newcommand{\jk}{\mbox{$J_K$}}

\newcommand{\nudu}{\mbox{$\nu_2$=1}}

\newcommand{\jkk}{\mbox{$J_{K_{\rm a},K_{\rm c}}$}}
\newcommand{\jkul}[4]{\mbox{$#1_{#2}$--$#3_{#4}$}}
\newcommand{\jkkul}[6]{\mbox{$#1_{#2,#3}$--$#4_{#5,#6}$}}
\newcommand{\jkkl}[3]{\mbox{$#1_{#2,#3}$}}
\newcommand{\twopihalf}{\mbox{$^{2}\Pi_{1/2}$}}

\newcommand{\jtmdm}{\mbox{$J$=$3/2$--$1/2$}}

\newcommand{\HtreceCN}{\mbox{H$^{13}$CN}}

\newcommand{\ammonia}{\mbox{NH$_3$}}
\newcommand{\vnsio}{\mbox{$^{29}$SiO}}
\newcommand{\trsio}{\mbox{$^{30}$SiO}}

\newcommand{\water}{\mbox{H$_2$O}}
\newcommand{\pwater}{\mbox{$p$-H$_2$O}}
\newcommand{\owater}{\mbox{$o$-H$_2$O}}
\newcommand{\pwaterdo}{\mbox{$p$-H$_2^{18}$O}}

\newcommand{\hydroxil}{\mbox{OH}}

\usepackage{txfonts}
\begin{document}
\title{\textit{Herschel}/HIFI observations of red supergiants and yellow hypergiants\thanks{Herschel is an ESA space observatory with 
science instruments 
provided by European-led Principal Investigator consortia and with 
important participation from NASA.},\thanks{Figures 10--13 and Appendix are only available in electronic form at http://www.edpsciences.org},\thanks{The spectra presented in this paper are available in electronic form at the CDS}}
\subtitle{I. Molecular inventory}
\author{D.~Teyssier\inst{1} 
\and G. Quintana-Lacaci\inst{2,5} 
\and A.P.~Marston\inst{1} 
\and V.~Bujarrabal\inst{3} 
\and J.~Alcolea\inst{4}
\and J.~Cernicharo\inst{5}
\and L.~Decin\inst{6,7}
\and C.~Dominik\inst{7,8}
\and K.~Justtanont\inst{9}
\and A.~de~Koter\inst{7,10}
\and G.~Melnick\inst{11}
\and K.M.~Menten\inst{12}
\and D.A.~Neufeld\inst{13}
\and H.~Olofsson\inst{9,14}
\and P.~Planesas\inst{4}
\and M.~Schmidt\inst{15}
\and R.~Soria-Ruiz\inst{4}
\and F.L.~Sch\"oier\inst{9}$^{,}$\thanks{Deceased 14 January 2011}
\and R.~Szczerba\inst{15}
\and L.B.F.M.~Waters\inst{7,16}
}

\institute{
European Space Astronomy Centre, Urb. Villafranca del Castillo, P.O. Box 50727, Madrid 28080, Spain 
\email{David.Teyssier@esa.int}
\and
Instituto de Radio Astronom\'ia Milim\'etrica (IRAM), Avenida Divina Pastora 7, Local 20, 18012 Granada, Spain
\and
Observatorio Astron\'omico Nacional (IGN), Ap 112, E--28803 Alcal\'a de Henares, Spain
\and
Observatorio Astron\'omico Nacional (IGN), Alfonso XII N$^{\circ}$3, E--28014 Madrid, Spain  
\and
CAB, INTA-CSIC, Ctra de Torrej\'on a Ajalvir, km 4, E--28850 Torrej\'on de Ardoz, Madrid, Spain
\and
Instituut voor Sterrenkunde, Katholieke Universiteit Leuven, Celestijnenlaan 200D, 3001Leuven, Belgium
\and
Sterrenkundig Instituut Anton Pannekoek, University of Amsterdam, Science Park 904, NL-1098 Amsterdam, The Netherlands
\and 
Department of Astrophysics/IMAPP, Radboud University Nijmegen, Nijmegen, The Netherlands
\and
Onsala Space Observatory,  Dept. of Radio and Space Science, Chalmers University of Technology, SE--43992 Onsala, Sweden
\and
Astronomical Institute, Utrecht University, Princetonplein 5, 3584 CC Utrecht, The Netherlands 
\and
Harvard-Smithsonian Center for Astrophysics, Cambridge, MA 02138, USA
\and
Max-Planck-Institut f{\"u}r Radioastronomie, Auf dem H{\"u}gel 69, D-53121 Bonn, Germany 
\and
The Johns Hopkins University, 3400 North Charles St, Baltimore, MD 21218, USA
\and
Department of Astronomy, AlbaNova University Center, Stockholm University, SE--10691 Stockholm, Sweden
\and
N. Copernicus Astronomical Center, Rabia{\'n}ska 8, 87-100 Toru{\'n}, Poland
\and
SRON Netherlands Institute for Space Research, Sorbonnelaan 2, 3584 CA Utrecht, The Netherlands
}

   \date{Received 04 May 2012 / Accepted 26 July 2012}

\authorrunning {D.~Teyssier et al.} 
\titlerunning{{\it Herschel}/HIFI observations of red supergiants and yellow hypergiants}

\abstract
{Red supergiant stars (RSGs) and yellow hypergiant stars (YHGs) are
believed to be the high-mass counterparts of stars in the asymptotic giant branch (AGB) and
early post-AGB phases. As such, they are scarcer and the properties and
evolution of their envelopes are still poorly understood. }
{We study the mass-loss in the post main-sequence evolution of
massive stars, through the properties of their envelopes in the 
intermediate and warm gas layers. These are the regions where the
acceleration of the gas takes place and the most recent mass-loss episodes
can~be~seen.}
{We used the HIFI instrument on-board the Herschel Space Observatory to observe sub-millimetre
and far-infrared (FIR) transitions of CO, water, and their isotopologues in a sample of two
RSGs (NML\,Cyg and Betelgeuse) and two YHGs 
(IRC+10420 and AFGL\,2343) stars. 
We present an inventory of the detected lines
and analyse the information revealed by their spectral profiles.
A comparison of the line intensity and shape in various
transitions is used to qualitatively derive a picture of the envelope
physical structure. On the basis of the results presented in an earlier study, we
model the CO and \treceCO\ emission in IRC+10420 and compare it to a set
of lines ranging from the millimetre to the FIR.}
{Red supergiants have stronger high-excitation lines
than the YHGs, indicating that they harbour dense and hot
inner shells contributing to these transitions. Consequently, these
high-$J$ lines in RSGs originate from acceleration layers that have not yet
reached the circumstellar terminal velocity and have narrower profiles
than their flat-topped lower-$J$ counterparts. The YHGs tend to
lack this inner component, in line with the picture of detached, hollow
envelopes derived from studies at longer wavelengths. NH$_3$ is only
detected in two sources (NML\,Cyg and IRC+10420), which are 
also observed to
be the strongest water-line emitters of the studied sample. In contrast, OH 
is detected in all sources and does not seem to correlate with
the water line intensities. We show that the IRC+10420 model derived
solely from millimetre low-$J$ CO transitions is capable of reproducing
the high-$J$ transitions when the temperature in the inner
shell is simply lowered by about 30\%.}
{}


\maketitle

\section{Introduction}
\label{intro}

Red supergiant stars (RSGs) and yellow hypergiant stars (YHGs) are
thought to be stages in the post main-sequence evolution of stars
with initial masses between $\sim$10 solar masses and 50 \ms\ (e.g.\ de Jager
1998, Meynet \& Maeder 2003, Levesque 2010). As such, RSGs and YHGs
are the massive counterparts of AGB and (early) post-AGB stars. However,
the very different properties of evolved massive stellar objects (RSGs, YHGs, 
Wolf-Rayet stars, luminous blue variable stars, Cepheid-like variables, 
supernovae etc), particularly their distribution in the H-R diagram, is hard to interpret with a simple
description of the evolution in this phase.

Both RSGs and YHGs are known to show complex mass-loss phenomena, which form
circumstellar envelopes (CSEs) that can be very dense. Mass-loss
plays an important role in the evolution of these objects
(de Jager 1998, Meynet \& Maeder 2003): almost one half of their total
initial mass can be ejected during these late phases, 
affecting in particular their possible later evolution into supernova events. 
Some objects have very high
mass-loss rates, with episodic rates as high as \mloss\ $\sim$ 10$^{-3}$ \my,
while the circumstellar envelopes around others are very diffuse,
corresponding to rates under 10$^{-7}$ \my\ (e.g. Castro-Carrizo et al.\
2007 - CC07 thereafter, Quintana-Lacaci 2008, Mauron \& Josselin 2011). From the
theoretical point of view, very high rates are also expected, at least
episodically, owing to a combination of high radiation pressure and
atmospheric activity (e.g.\ Josselin \& Plez 2007; as
happens in the AGB) or to the intrinsic instabilities characteristic of
the yellow phases of the late evolution of high-mass stars (the so-called
"yellow void"; see e.g.\ de Jager 1998).

Molecular lines have been one of the most
powerful tools for the study of the properties of CSEs around AGB and post-AGB
stars. Similar data on RSGs and YHGs are rare, but in some cases
molecular lines have yielded a quite comprehensive insight into their CSEs. 
For instance, using model fitting of mm- and
submm-observations of several CO rotational lines in the RSG VY CMa,
Decin et al.\ (2006) deduced mass-loss rates as high as 3$\times$10$^{-4}$ \my, that varied on timescales of
about 1000 yr.  VY CMa is known to harbour a wide variety of molecular
species, illustrating the very complex chemistry at play in its CSE
(e.g.\ Tenenbaum et al.\ 2010 and references therein). Other less well-studied RSGs, such as
VX Sgr and NML Cyg, also show high mass-loss rates of the order of $\sim$
10$^{-4}$ \my (e.g.\ De Beck et al.\ 2010).
An example of a low-\mloss\ RSG is Betelgeuse
($\alpha$\,Ori), with a value of about 10$^{-7}$-10$^{-6}$ \my, although this
value is uncertain because its molecular emission may be weaker than in
other objects owing to its molecular underabundance (see Huggins et al.\ 1994, Mauron \& Josselin
2011).  The YHGs IRC+10420 and AFGL\,2343 have also
been well-studied in molecular emission. 
CC07 (see also Quintana-Lacaci et al.\ 2008) compared line-emission 
models with high-resolution maps of the CO \juc\ and \jdu\
lines, deriving large \mloss\ variations on timescales of about
1000 yr and \mloss\ maxima in excess of 10$^{-3}$ \my. These objects also have
many molecular lines and a very rich chemistry (Quintana-Lacaci et al.\
2007). The total masses in the molecule-rich shells are particularly
high for these two YHGs, $M_{\rm tot}$ \gsim\ 1 \ms. The total masses
 of the molecular envelopes around RSGs are somewhat lower,
between 0.1 solar masses and 1 \ms\ for VY CMa, VX Sgr, and NML Cyg.

Despite the observational progress that has been made to date,
 the available data still fail to provide information on
some basic parameters. Low-$J$ CO lines are good tracers of both the
circumstellar mass distribution and kinematics, but these easily
excited lines cannot probe warm gas with temperatures $T_{\rm k}$
\gsim\ 100 K, which are expected to be present and sometimes
dominant in these shells. The temperature itself is not well-determined 
under these conditions, which may affect the determination
of the mass-loss rate and the total mass. To properly study these warm
components, it is therefore necessary to observe lines in the far-infrared (FIR),
involving level excitations comparable (in temperature units) to these
moderate kinetic temperatures.

This paper presents {\it Herschel}/HIFI observations in
the FIR of molecular lines from two RSGs, Betelgeuse and NML Cyg, and
two YHGs, IRC+10420 and AFGL\,2343. The observations are part of the
Herschel guaranteed time key program HIFISTARS (PI V. Bujarrabal), devoted to the study
of high-excitation molecular lines in (low- and high-mass) evolved
stars. Companion observations of VY CMa, which are particularly rich and
complex, will be discussed in another paper (Alcolea et
al., in prep.). We present data of the \jsc, \jdn, and \jdsq\ lines of
\doce\ and \trece, which are good probes of the excitation.
We also discuss our observations of other molecules, including water vapour, OH, and NH$_3$, 
which are particularly useful to understanding more clearly the chemistry in these unusual objects.

\section{Observations and data processing}
\label{obs}

The observations presented in this paper were obtained with the
Heterodyne Instrument for the Far Infrared (thereafter HIFI, de Graauw
et al.~2010), on-board the Herschel Space Observatory (Pilbratt et
al.~2011). Similarly to other observations presented in the framework
of the HIFISTARS key program (e.g. Bujarrabal et al.~2012), the main
target lines of this project were collected through a handful of
HIFI settings. Thanks to the instantaneous spectral coverage of 2.4 - 4 GHz 
offered by the Wide-Band Spectrometer (WBS), and since
HIFI works with double-side-band (DSB) mixers, several other bonus
lines could be covered simultaneously. A summary of the settings used
and their main characteristics is given in Table~\ref{tabletel}.

All data were taken in the double-beam switching (DBS) mode,
consisting of chopping and nodding between the source line-of-sight and
blank sky positions 3~arcmin on either sides of the source. This mode
allows very fast
switches between the sky positions, and therefore mitigates the
detector drifts. Residual baseline ripples can however remain,
especially in the HIFI bands 6 and 7 where the detectors used (hot
electron bolometers, hereafter HEB) are more susceptible to
instabilities.

We used the standard pipeline products that had been generated and made
available at the Herschel Science Archive (HSA). These could be accessed via 
the HIPE software\footnote{HIPE is a joint development by the Herschel Science Ground Segment
Consortium, consisting of ESA, the NASA Herschel Science Center, and
the HIFI, PACS, and SPIRE consortia.}. The HIFI pipeline brings all data onto a
single-side-band $T_{\rm A}^*$ intensity scale. 
A side-band ratio of unity was assumed for all the lines reported in this paper.
All individual spectra from the so-called {\it level1} were checked for outliers, and averaged
for each of the two orthogonal mixer polarisations available on
HIFI. In some rare cases, one of the polarisations was useless owing to
an unpumped mixer, so its data had to be discarded. The sky positions
observed by the two polarization mixers are not strictly the same, the
misalignment varying between $\sim$ 6\arcsec~in the lower bands to
1\arcsec~in the higher bands (Roelfsema et al.~2012). Any such misalignment could 
in practice manifest itself (among other calibration effects) as intensity imbalance
between the two polarisations. We however averaged the two spectra whenever 
possible to improve the sensitivity. In practice, this implies that the effective beam size of the
averaged data is slightly larger than the nominal spatial
resolution.

\begin{table}
\caption{Summary of observational settings and telescope characteristics. The setting names are provided for cross-reference with other HIFISTARS results, and to ease the association of data with a given Herschel observation identifier ({\it obsid}). These obsid's are given in the figures of the appendix. For each setting, we indicate the corresponding HIFI band, the local oscillator frequency tuned for the observation and the system noise temperature achieved on average, together with the typical beam properties applying to each of them.}             
\label{tabletel}      
\centering          
\begin{tabular}{cccccc}     
\hline\hline       
Setting & LO freq. & Typical $T_{\rm sys}$ & Band & Beam      & $B_{\rm eff}$        \\
 name      & (GHz)        &      (K, DSB)    &      & size      &       \\
\hline
14g     &  ~~564.56    &    ~~~~93     & 1B   & 37.5\arcsec &   0.754    \\
12f     &   ~~653.55    &     ~~131     & 2A   & 32.4\arcsec &   0.752    \\
17     &   ~~686.42    &     ~~142     & 2A   & 30.9\arcsec &   0.745    \\
07e       & 1106.90    &     ~~416     & 4B   & 19.1\arcsec &   0.742    \\
06d      & 1157.67    &     ~~900     & 5A   & 18.3\arcsec &   0.639     \\
05c        & 1200.90    &      1015     & 5A   & 17.6\arcsec &   0.633    \\
03b        & 1757.68    &      1580     & 7A   & 12.1\arcsec &   0.740    \\
16       &  1838.31    &      1300     & 7B   & 11.5\arcsec &   0.736    \\
\hline\hline
\end{tabular}
\end{table}

A special treatment had to be applied to some of the data taken in the
HEB bands owing to the standing wave affecting in
particular these detectors. This spectral artefact is not optical in nature, hence 
cannot be optimally treated by applying the standard defringing
methods to sine-wave baseline distortion. A novel method is
currently being developed to perform an alternative pipeline correction
of these features (Higgins~2011), which may, however, be ineffective when
the line widths become similar to those of the artefact
structure. Because this particular situation applies to the sources treated in
this study, we decided not to adopt this approach. We developed instead a
semi-automated procedure that discards individual
spectra where the contribution of the baseline ripple is significantly stronger
than the expected radiometric noise. This results in the rejection of
up to half of the non-averaged spectra, and therefore increases the
noise in the final data, in spite of the more reliable baseline.

Finally, all spectra were converted onto a $T_{\rm mb}$ scale, by applying
the main beam efficiencies reported by the HIFI Instrument Control Centre, which are reproduced in
Table~\ref{tabletel}. 
We based our absolute-calibration accuracy estimates
on the error budget reported by Roelfsema et al. (2012). On top of this, we
considered an additional contribution arising from the standing wave 
described above for bands 6 and 7. Adding up all error contributions in quadrature
only provided lower limits to the typical calibration errors.
In practice, the calibration error that we assumed in this study is 15\% in bands 1 and 2, 20\% in
bands 3 to 5, and 30\% in bands 6 and 7.  The collection of all full-band spectra is compiled in
Appendix~\ref{appendix}.

\section{Results}
\label{result}

\begin{table*}
\caption{Observed coordinates and additional stellar
parameters for the two RSGs and the two YHGs considered in this study: spectral type, distance and luminosity. References are: 1: Mauron \& Josselin (2011); 
2: Schuster et al.\ (2009);  3: de Jager (1998); 4: Hawkins et al.\
(1995), Reddy \& Hrivnak (1999).}
\label{tablesou}
\begin{center}
\begin{tabular}{lcccccl}
\hline
\hline
Source & $\alpha$ (J2000) & $\delta$ (J2000) & Spectral type & Distance (pc) &
Luminosity (\ls) & References \\
\hline
\multicolumn{5}{l}{\bf Red supergiants} \\
\hline
Betelgeuse & 05 55 10.3  & +07 24 25.4 & M2 & 150  & 4 10$^5$ & 1 \\
NML\,Cyg   & 20 46 25.5  & +40 06 59.6 & M6 & 1700 & 6 10$^5$ & 2 \\ 
\hline
\hline
\multicolumn{5}{l}{\bf Yellow hypergiants} \\
\hline
IRC+10420 & 19 26 48.0 & +11 21 16.7 & F8 & 5000 & 7 10$^5$ & 3 \\
AFGL\,2343 & 19 13 58.6 & +00 07 31.9 & G5 & 6000 & 6 10$^5$ & 4 \\
\hline
\end{tabular}
\end{center}
\end{table*}

\begin{table*}
\caption{Spectral line results for NML\,Cyg and IRC+10420. For each line we indicate the corresponding upper energy level and rest frequency, together with the measured line peak intensity, line integrated intensity, and the velocity range used to compute this latter.}             
\label{tablines1}      
\centering          
\begin{tabular}{rlcccccccc}     
\hline\hline 
          &            &    & &  \multicolumn{3}{c}{NML\,Cyg}   & \multicolumn{3}{c}{IRC+10420}  \\
          &            & $E_{\rm upp}^\dagger$  & Rest freq.  & Peak$^{(7)}$ & Integ. intensity   & Vel. range$^{(2)}$  & Peak$^{(7)}$ & Integ. intensity   & Vel. range$^{(2)}$     \\
Species   & Transition &  (K) &  (GHz)     &  (mK)  & (K\,\kms)& LSR (\kms)  & (mK) & (K\,\kms)& LSR (\kms)   \\
\hline
\doceCO                             & \jsc                           & 116& 691.473   &    471({\it 13})$^{(4)}$ &  19.6 & [--34; 33]    &  364({\it 20})$^{(3)}$ &  19.1 & [37; 115] \\
                                    & \jdn                           & 304&  1151.985    &  479({\it   39})$^{(4)}$ &  19.1 & [--33; 30]   &   243({\it   44})$^{(4)}$ &  14.7 & [36; 114]     \\
                                    & \jdsq                          & 752&  1841.346     &  666({\it   47})$^{(4)}$ &  17.6 & [--38; 28]  &     133({\it   44})$^{(5)}$ &  6.57 & [38; 111] \\
\\                                                                                                                                                               
\treceCO                            & \jsc                           & 111& 661.067   &  95({\it   7})$^{(4)}$ &  4.06 & [--34; 33]  &   57({\it   6})$^{(4)}$ &  2.79 & [36; 113] \\
                                    & \jdn                           & 291&  1101.350    &  141({\it   17})$^{(4)}$ &  3.80 & [--25; 31]      &  60({\it   20})$^{(4)}$ &  2.8 & [43; 107] \\
                                    & \jdsq                          & 719&  1760.486     & 135({\it   43})$^{(6)}$ &  4.75 & [--31; 27]    &  $<$138$^{(1,5)}$  & --  & --  \\
\\                                                                                                                                                               
\owater                             & \jkk=\jkkul{1}{1}{0}{1}{0}{1}  &27& 556.936  &  386({\it  6})$^{(3)}$ &  12.2 & [--23; 37]    &  168({\it 6})$^{(3)}$ &  6.80 & [43; 118] \\
                                    & \jkk=\jkkul{3}{1}{2}{2}{2}{1}  & 215&  1153.127  &   848({\it   41})$^{(4)}$ &  28.8 & [--23; 35]    &  333({\it   42})$^{(4)}$ &  13.7 & [38; 111]  \\
                                    & \jkk=\jkkul{3}{2}{1}{3}{1}{2}  & 271&  1162.912   &     869({\it   42})$^{(4)}$ &  29.7 & [--31; 34]     &  195({\it   41})$^{(4)}$ &  8.41 & [40 ;112]  \\
\\                                                                                                                                                               
\pwater                             & \jkk=\jkkul{1}{1}{1}{0}{0}{0}  &53&  1113.343       &  993({\it   24})$^{(3)}$ &  30.8& [--23; 38]     &  329({\it   21})$^{(3)}$ &  12.7 & [46; 115]  \\
                                    & \jkk=\jkkul{4}{2}{2}{4}{1}{3}  & 454&  1207.639   &   726({\it   47})$^{(5)}$ &  22.9 & [--32; 34]    &  $<$144$^{(1,4)}$ &   {\bf --} & {\bf --}  \\
                                    & \jkk=\jkkul{6}{3}{3}{6}{2}{4}  & 952&  1762.043     &   526({\it   43})$^{(4)}$ &  14.0 & [--29; 30]     &    $<$180$^{(1,4)}$  &    --    &    --  \\
\\                                                                                                                                                               
\owater\ \nudu                      & \jkk=\jkkul{1}{1}{0}{1}{0}{1}  & 2326 &  658.004 & 882({\it   9})$^{(3)}$ &  9.53 & [--26; 18]  &    $<$15$^{(1,5)}$  &    --    &    --  \\
\\                                                                                                                                                               
\pwater\ \nudu                      & \jkk=\jkkul{1}{1}{1}{0}{0}{0}  & 2352 & 1205.788    & 112({\it  44})$^{(4)}$ &  2.64 & [--18; 17]      &    $<$144$^{(1,4)}$  &    --    &    --  \\
\\                                                                                                                                                               
\pwaterdo                           & \jkk=\jkkul{1}{1}{1}{0}{0}{0}  &53&  1101.698       & 185({\it   18})$^{(4)}$ &  5.05 & [--24; 37]     &    $<$60$^{(1,4)}$  &    --    &    --  \\
\\
SiO                                    & $J$=16--15                     &283 & 694.294    &  118({\it   14})$^{(4)}$ &  3.42 & [--32; 30]       &    $<$35$^{(1,5)}$  &    --    &    --  \\
\\                                                                                                                                                               
SiO v=1                             & $J$=13--12                     & 1957 & 560.325     &  17({\it   3})$^{(5)}$ &  0.39 & [--25; 27] &    $<$15$^{(1,4)}$  &    --    &    --  \\
                                    & $J$=15--14                     & 2017 & 646.429   &  30({\it   5})$^{(5)}$ &  0.50& [--25; 25] &    $<$15$^{(1,5)}$  &    --    &    --  \\
\\
\vnsio                              & $J$=13--12                     &187 & 557.179    & 30({\it   4})$^{(5)}$ &  0.95 & [--25; 26]      &    $<$15$^{(1,4)}$  &    --    &    --  \\
                                    & $J$=26--25                     &722 &  1112.798  &    50({\it   17})$^{(5)}$ &  1.10 & [--21; 13]    &    $<$60$^{(1,4)}$  &    --    &    --  \\
\\                                                                                                                                                               
\trsio                              & $J$=26--25                     &713 &  1099.708  & 59({\it   14})$^{(5)}$ &  0.77 & [--17; 24]    &  $<$60$^{(1,4)}$  &    --    &    --    \\
\\                                                                                                                                                               
SO                                  & \jk=\jkul{13}{14}{12}{13}      &193 & 560.178     & 13({\it  4})$^{(5)}$  &  0.40 & [--22; 23]     &    $<$15$^{(1,4)}$  &    --    &    --  \\
                                    & \jk=\jkul{13}{12}{12}{11}      &194 & 558.087      & 13({\it   4})$^{(5)}$  &  0.56 & [--35; 29]    &    $<$15$^{(1,4)}$  &    --    &    --  \\
                                    & \jk=\jkul{13}{13}{12}{12}      &201 & 559.320       & 20({\it   3})$^{(5)}$  &  0.49 & [--22; 25]    &    $<$15$^{(1,4)}$  &    --    &    --  \\
                                    & \jk=\jkul{15}{16}{14}{15}      &253 & 645.875      & 21({\it   4})$^{(5)}$  &  0.54 & [--21; 22]      &    $<$15$^{(1,5)}$  &    --    &    --  \\
\\                                                                                                                                                               
H$_2$S		        & \jkk=\jkkul{3}{1}{2}{2}{2}{1}  & 137&  1196.012  & 149({\it 33})$^{(4)}$ &  3.83 & [--13; 22]      &    $<$144$^{(1,4)}$  &    --    &    --  \\
\\
HCN                                    & $J$=13--12                     & 386&  1151.452      &  86({\it   27})$^{(6)}$ &  3.48 & [--35; 32]       &    $<$126$^{(1,4)}$  &    --    &    --  \\
\\                                                                                                                                                            
\HtreceCN                           & $J$=8--7                       &147& 690.552      & $<$18$^{(1,6)}$  &  {\bf --} & {\bf --}  &    $<$35$^{(1,5)}$  &    --    &    --  \\
\\                                                                                                                                                               
\ammonia                            & \jk=\jkul{1}{0}{0}{0}          &27& 572.498     & 102({\it   3})$^{(4)}$ &  4.08 & [--38; 38]    &  93({\it  5})$^{(4)}$ &  2.93 & [41; 116] \\
\\                                                                                                                                                               
\hydroxil\ \twopihalf               & \jtmdm                         &270 &  1834.747  &    800({\it   46})$^{(4)}$ &  31.4 & [--25; 36]    &  599({\it  58})$^{(4)}$ &  33.5 & [37; 113] \\                                                                                                                        
\hline\hline
\multicolumn{10}{l}{$^\dagger$ For \owater, the energies are relative to that of the lowest ortho level (\jkk=\jkkl{1}{0}{1}), since 
\owater\ and \pwater\ behave as different species.} \\
\multicolumn{10}{l}{$^{(1)}$ Non-detections are given as 3-$\sigma$, $^{(2)}$ Range over which the integrated intensity is computed, $^{(3)}$ Computed in bins of order 0.5\,\kms,} \\
\multicolumn{10}{l}{$^{(4)}$ Computed in bins of order 1\,\kms, $^{(5)}$ Computed in bins of order 2\,\kms, $^{(6)}$ Computed in bins of order 3\,\kms,} \\
\multicolumn{10}{l}{$^{(7)}$ Computed as the line maximum minus 1.5$\times$ the noise rms, indicated in italics between brackets (1-$\sigma$)}\\
\end{tabular}
\end{table*}    

\begin{table*}
\caption{Same as Table~\ref{tablines1} for the spectral line observed in AFGL\,2343 and Betelgeuse}             
\label{tablines2}      
\centering          
\begin{tabular}{rlcccccccc}     
\hline\hline 
          &            &    & &  \multicolumn{3}{c}{Betelgeuse}   & \multicolumn{3}{c}{AFGL\,2343}  \\
          &            & $E_{\rm upp}^\dagger$  & Rest freq.  & Peak$^{(6)}$ & Integ. intensity   & Vel. range$^{(2)}$  & Peak$^{(6)}$ & Integ. intensity   & Vel. range$^{(2)}$     \\
Species   & Transition &  (K) &  (GHz)     &  (mK)  & (K\,\kms)& LSR (\kms)  & (mK) & (K\,\kms)& LSR (\kms)   \\
\hline
\doceCO                             & \jsc                           & 116& 691.473   &    654({\it 17})$^{(3)}$ &  13.2 & [--12; 24]    &  261({\it 13})$^{(4)}$ &  11.9 & [61; 135] \\  
                                    & \jdn                           & 304&  1151.985    &  762({\it 54})$^{(3)}$ &  13.4 & [--12; 21]     &   211({\it 26})$^{(5)}$ &  6.94 & [61; 137]   \\
                                    & \jdsq                          & 752&  1841.346     &  866({\it 58})$^{(3)}$ &  13.6 & [--11; 22]  &    $<$~111$^{(1,5)}$  &    --    &    --  \\
\\                                                                                                                                                               
\treceCO                            & \jsc                           & 111& 661.067   & 136({\it 7})$^{(3)}$ &  2.83 & [--10; 18]  &   87({\it 6})$^{(4)}$ &  3.00 & [62; 133] \\
                                    & \jdn                           & 291&  1101.350    &  160({\it   20})$^{(3)}$ &  2.45 & [--9; 23]       &  $<$~156$^{(1,4)}$  &    --    &    --   \\
                                    & \jdsq                          & 719&  1760.486     & 187({\it 50})$^{(5)}$ &  2.68 & [--9; 22]     &  $<$~105$^{(1,5)}$  &    --    &    --    \\
\\                                                                                                                                                               
\owater                             & \jkk=\jkkul{1}{1}{0}{1}{0}{1}  &27& 556.936  &  35({\it 4})$^{(4)}$ & 0.61 & [--6; 21]      &  36({\it 4})$^{(4)}$ &  1.41 & [62; 132]  \\
                                         & \jkk=\jkkul{3}{2}{1}{3}{1}{2}  & 271&  1162.912   &  $<$~112$^{(1,5)}$   &  -- & --    &  98({\it   35})$^{(5)}$ &  4.26 & [73 ;131]  \\
\\                                                                                                                                                               
\pwater                             & \jkk=\jkkul{1}{1}{1}{0}{0}{0}  &53&  1113.343  &  72({\it 16})$^{(4)}$ &  1.54 & [--9; 21]     &  49({\it 13})$^{(4)}$ &  1.93 & [77; 130]  \\
\\                                                                                                                                                               
\hydroxil\ \twopihalf               & \jtmdm                         &270 &  1834.747  &     208({\it 34})$^{(4)}$ &  4.45 & [--10; 20]     &  362({\it 30})$^{(5)}$ &  14.0 & [64; 129]   \\
\hline\hline
\multicolumn{10}{l}{$^\dagger$ For \owater, the energies are relative to that of the lowest ortho level (\jkk=\jkkl{1}{0}{1}), since 
\owater\ and \pwater\ behave as different species.} \\
\multicolumn{10}{l}{$^{(1)}$ Non-detections are given as 3-$\sigma$, $^{(2)}$ range over which the integrated intensity is computed, $^{(3)}$ Computed in bins of order 0.5\,\kms,} \\
\multicolumn{10}{l}{$^{(4)}$ Computed in bins of order 1\,\kms, $^{(5)}$ Computed in bins of order 2\,\kms,} \\
\multicolumn{10}{l}{ $^{(6)}$ Computed as the line maximum minus 1.5$\times$ the noise rms, indicated in italics between brackets (1-$\sigma$)}\\

\end{tabular}
\end{table*}    

Table~\ref{tablesou} summarises the main stellar parameters of the
sample studied here. In the following sub-sections, we give an overall
inventory of all probed and detected species, together with the
extracted line characteristics in the form of the line peak 
and integrated intensities (see Tables~\ref{tablines1} and~\ref{tablines2} for a
compilation of these results).
The high spectral resolution of HIFI also allows us to distinguish and discuss the
contribution of particular shell layers to the emission
of the respective species.

\subsection{NML Cyg}
\label{nmlcyg}

\begin{figure}
\centering
\includegraphics[angle=0.0,width=\columnwidth]{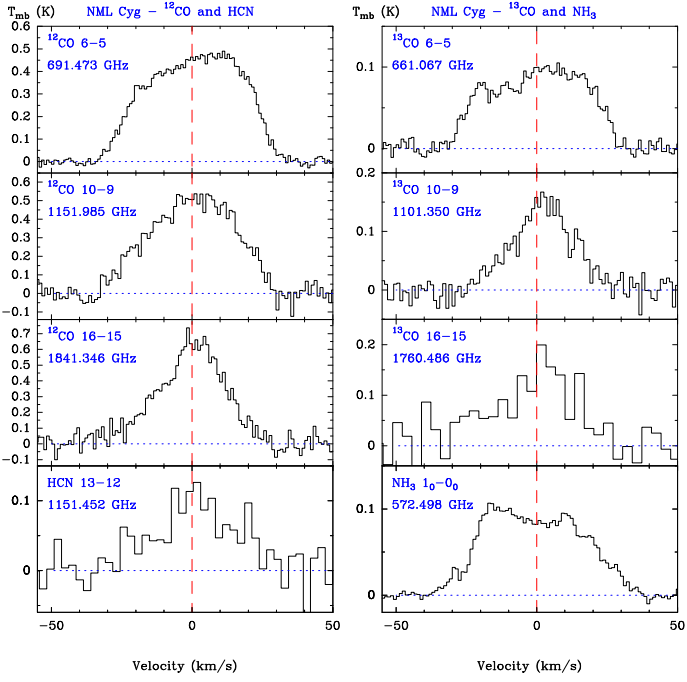}\\
\caption{Individual line spectra collected with {\it Herschel}/HIFI on NML\,Cyg for CO, \treceCO, NH$_3$, and HCN.
The dashed line indicates the source systemic velocity. }
\label{fig_nml_co}
\end{figure}

\begin{figure}
\centering
\includegraphics[angle=0.0,width=\columnwidth]{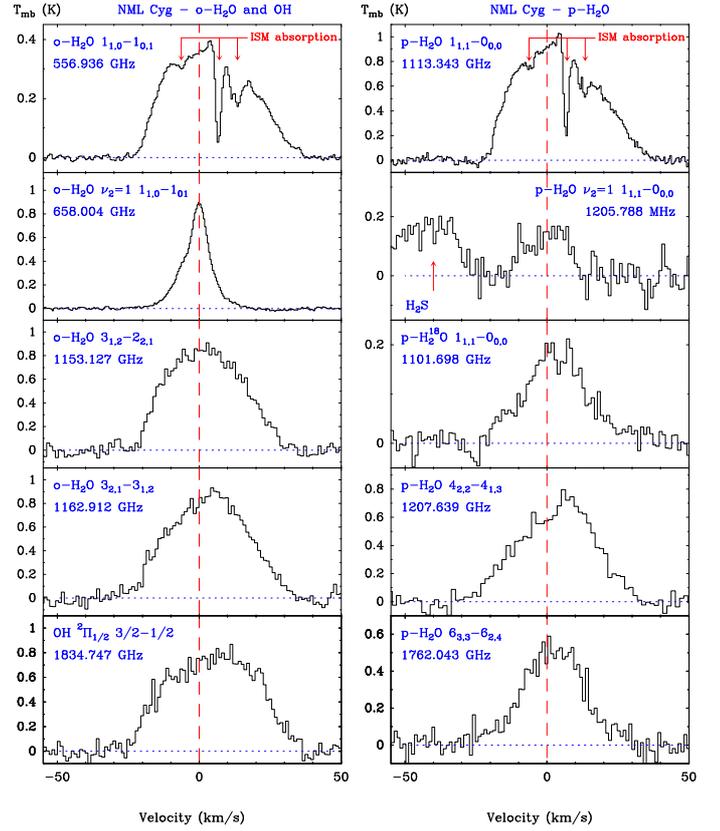}\\
\caption{Same as Fig.~\ref{fig_nml_co} for OH, \water\ and isotopologues in NML\,Cyg}
\label{fig_nml_h2o}
\end{figure}

\begin{figure}
\centering
\includegraphics[angle=0.0,width=\columnwidth]{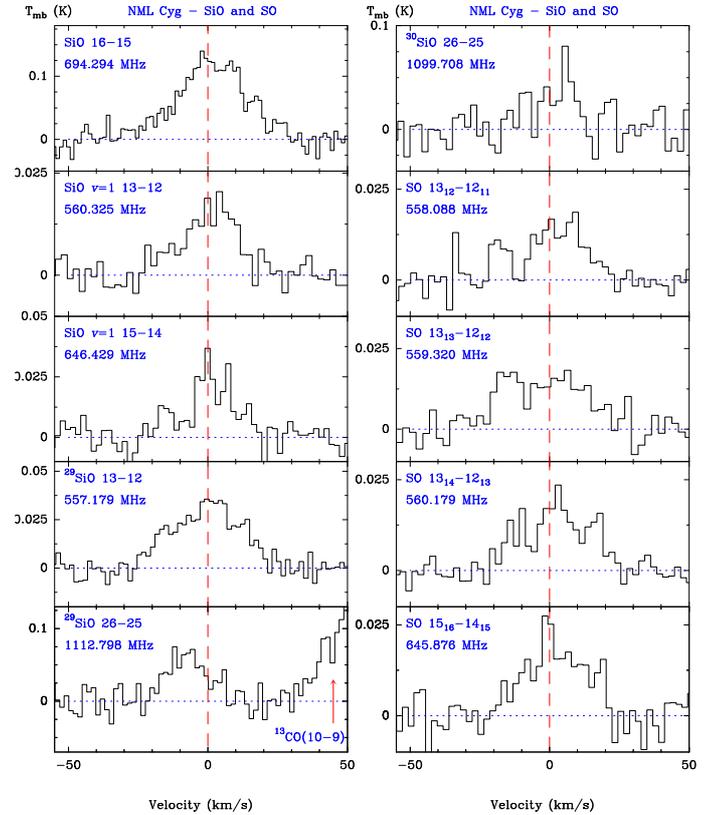}\\
\caption{Same as Fig.~\ref{fig_nml_h2o} for SiO, $^{29}$SiO, $^{30}$SiO and SO in NML\,Cyg.}
\label{fig_nml_sio}
\end{figure}

NML\,Cyg is a red supergiant of spectral type M6.  It has an
intermediate mass-loss rate of order 8.7$\times$10$^{-5}$\my (De Beck
et al. 2010).

NML\,Cyg exhibits the richest spectrum of all four sources presented
here, although, among the full sample of RSGs, it is largely surpassed by the line density of VY\,CMa
(Alcolea et al., in prep.). Apart from this latter, it is also
the source with the strongest line emission when considering species
and transitions individually (see Tables~\ref{tablines1} and~\ref{tablines2}, and
Fig.~\ref{fig_histo}). While the detection of water vapour lines in other sources is mostly
limited to the ground-state transitions of both {\it ortho}- and {\it
para}-\water, NML\,Cyg shows
transitions with energy levels in excess of 2000\,K. One noticeable
detection is that of the \water\ maser line at 658 GHz.
The measured flux of $\sim$400\,Jy agrees well with the first detection of this line 
by Menten \& Young (1995), indicating that there is no evidence of large variability.
Water maser emission at 22\,GHz has also been reported in this source
(e.g. Nagayama et al. 2008). While the velocity distribution of this
latter exhibits two shifted components at -25\,\kms\ and +5\,\kms\
respectively, the 658\,GHz maser is peaked at the star velocity
of v$_{\rm LSR}$ = 0\,\kms, with a relatively narrow profile
compared to the other submm transitions, and a blue-shifted shoulder. 
This is in contrast to the rest of the water vapour lines, where the blue side is usually narrower than the red side.
The narrower line width shows that the maser originates from inner layers still in acceleration.

NML\,Cyg also displays a uniquely rich spectrum of SO, rotational transitions from the ground, and 
vibrationally excited states of SiO, as well as isotopes of the SiO, though  
only from the ground vibrational state (see Fig.~\ref{fig_nml_sio}). 
The detection of high-$J$ transitions of SiO
confirms the presence of warm and dense gas at the centre of the
envelope. This is a noticeable observation since, as shown in the
following sections, the other sources seem to have
rather hollow envelopes.
This result is corroborated by the presence of very high
excitation \water\ transitions, and tells us that the
mass-loss rate is probably still very high today. Finally, NML\,Cyg is
one of the few sources showing a strong emission in \ammonia,
and the only RSG together with VY\,CMa to exhibit detectable emission from other
hydrides such as HCN and H$_2$S.

The velocity structure of the various species and transitions displays a
great variety of profile shapes. 
As is well-known, the line shape essentially
depends on the line opacity, but also on the velocity profile of the
layers where the line is emitting. Typically, the transition from a rectangular
(flat-top) to a parabolic profile reflects the effect of optical depth in a constant
velocity flow, while the transition to a triangular, narrower profile comes about
when one probes the inner and warmer acceleration region with higher excitation
lines. A recent study of submm molecular lines of CO and \water\ in O-rich AGB stars
(Justtanont et al. 2012) indicates that the line width tends to be smaller for high
excitation lines formed closer to the star, pointing to an accelerated outflow.

In NML\,Cyg, this kind of transition is particularly pronounced when looking at the
evolution between the \jsc, \jdn, and \jdsq\ transitions of \doceCO\ and \treceCO, respectively 
(see Fig.~\ref{fig_nml_co}). That the same trend is observed in both optically
thick and thin isotopes suggests that this is not primarily an opacity effect and we 
indeed detect the velocity gradient in the innermost layers of the envelope. 
The same behaviour is observed for increasingly higher energy levels of 
the \pwater\ and \pwaterdo\ transitions (see Fig.~\ref{fig_nml_h2o}). 
Opacity, however, also plays a role in the line shape, as can be seen from the narrowing of the
line profile between the respective \jdn\ transitions of \doceCO\ and \treceCO , which have similar
excitation temperatures.
Interestingly, the 1834 GHz OH line profile closely resembles that of the mid-$J$ CO 
transitions, suggesting that its emission arises predominantly in the outer layers where
the terminal velocity has been reached.

Another peculiarity of the velocity structure can be observed in the
\ammonia\ line profile (see Fig.~\ref{fig_nml_co}). 
Unlike the other lines studied, that exhibit an overall single-peaked profile, the
\ammonia\ emission has two separate peaks on either side of the systemic velocity.
In particular, this structure provides evidence for a component lying at velocity around -18\,\kms.
We note that this blue-shifted velocity component coincides with a prominent contribution of the 
$\nu$=1, \jdu, SiO maser emission at 43\,GHz (Boboltz \& Marvel 2000), showing up at sub-arcsec scales as a marked
bi-polar structure also reported in mid-IR maps of the dust emission (e.g. Monnier et al. 2004).
Menten et al. (2010) already emphasized the peculiar behaviour of ammonia in other giant stars
such as VY\,CMa or IRC+10420, reporting unexpectedly high abundances in their
envelope. We speculate that the bulk of the \ammonia\ abundance 
could actually be quite widely distributed over the circumstellar shell.

Finally, we note the various ISM absorption signatures
seen only in the two ground-state transitions of \owater\ and \pwater. 
While these features were also observed in low-$J$ CO
transitions (Kemper et al., 2003), they are not present in
the other transitions probed by {\it Herschel}, confirming that they arise from
either a diffuse or cold gas component.
We also note that there is a narrow emission component in
the ground state \owater\ and \pwater\ at 5\,\kms.
This small peak was also observed in the
\doceCO\ \jdu\ line of Kemper et al. (2003) and probably originates from a
small ISM clump along the line-of-sight. 
This interpretation is supported by recent distance measurements of NML\,Cyg
(Zhang et al. 2012) and foreground interstellar gas (Rygl et al. 2012).

\subsection{Betelgeuse}
\label{bet}

\begin{figure}
\centering
\includegraphics[angle=0.0,width=\columnwidth]{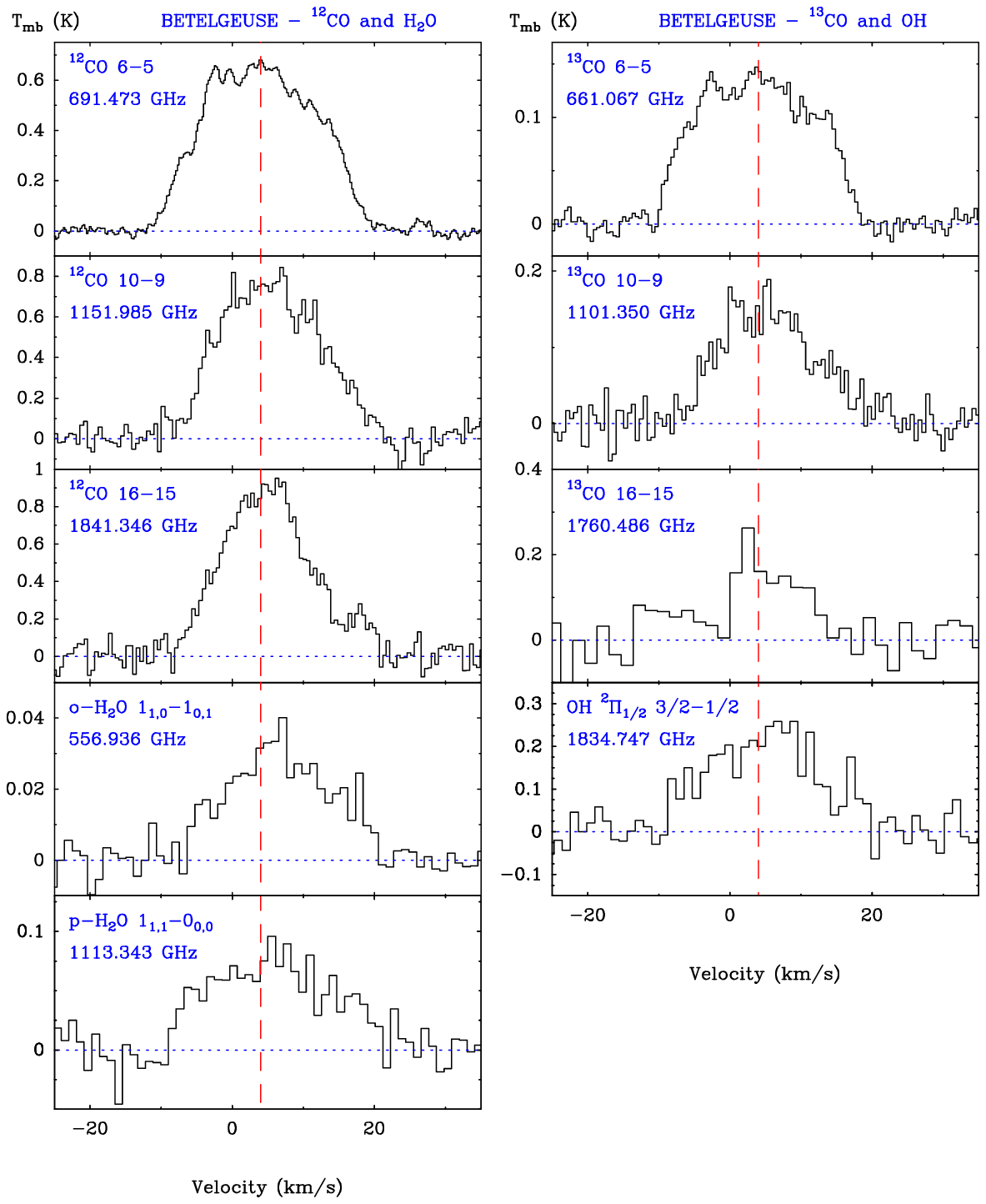}\\
\caption{Same as Fig.~\ref{fig_nml_h2o} for lines detected in Betelgeuse.}
\label{fig_bet}
\end{figure}

Betelgeuse ($\alpha$ Ori) is a red supergiant of spectral type M2. It
has a relatively low mass-loss rate of order 2$\times$10$^{-7}$\my (De
Beck et al. 2010).

Betelgeuse appears as a source that is particularly strong in \doceCO\ and
\treceCO\ emission. In absolute intensities, it is the brightest of all
targets reported here for these molecules. 
In contrast, it exhibits faint \water\ and OH lines, especially in relation to the CO line intensity 
(Fig.~\ref{fig_bet}, see also Fig.~\ref{fig_histo}). Betelgeuse has a
particularly small distance, 150 pc, which, despite its low mass-loss
rate, explains its high CO line intensities. Owing to this weak
mass loss, the envelope density is lower, and the weak water emission
could be a consequence of enhanced photo-dissociation by stellar and
interstellar ultra-violet (UV) photons as \water\ is
more easily photodissociated than CO (e.g. Justtanont et al. 1999). 
We recall that this source has very strong fine-structure
lines of both neutral atoms and low-energy ions (e.g.\ Haas \& Glassgold 1993, Castro-Carrizo et
al.\ 2001), which are thought to be abundant in
the inner envelope owing to the UV emission of the extended chromosphere
of the star.

As in NML Cyg, the \doceCO\ and \treceCO\ line profiles become 
narrower as the excitation temperature increases, which again suggests that the
high-transition lines form predominantly in the acceleration region.
Betelgeuse is thought to have a particularly hot inner circumstellar region (Rodgers \&
Glassgold 1991), with temperatures higher than about 1000 K out to
10$^{15}$ cm from the star (i.e.\ occupying about 1$''$).
The evolution in the line profile can be clearly seen if one compares the spectra
reported here to the nearly flat-top shape of the \doceCO\ \jtd\ spectrum observed by Kemper et al. (2003).
There are also some small spectral components in our CO profiles sitting on
top of the overall parabolic or triangular profiles. Those small
structures could be related to clumps in the inner envelope,  
which have been reported in mid-IR observations of the inner 3\arcsec of the nebula (Kervella et al. 2011).

\subsection{IRC+10420}
\label{irc}

\begin{figure}
\centering
\includegraphics[angle=0.0,width=\columnwidth]{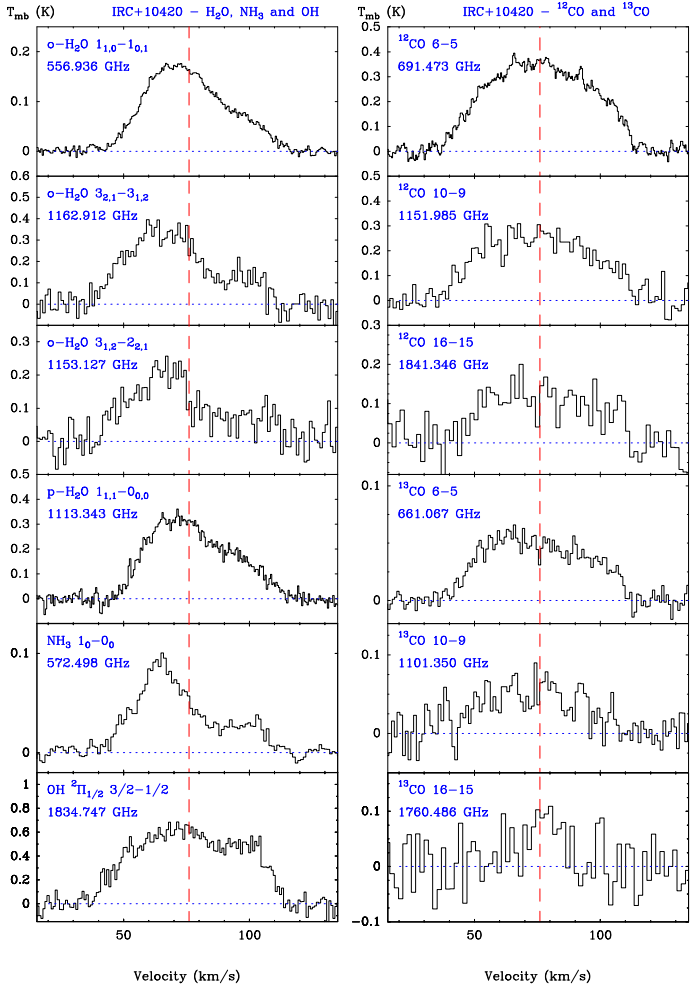}
\caption{Same as Fig.~\ref{fig_nml_h2o} for lines detected in IRC+10420.}
\label{fig_irc}
\end{figure}

IRC+10420 is a yellow hypergiant, of spectral type F8. It has a
strongly variable mass-loss rate, with highest values around 
3$\times$10$^{-4}$\my (CC07).

IRC+10420 presents the second richest species inventory after NML\,Cyg (Fig.~\ref{fig_irc}),
although it features none of the SO, SiO, or C-bearing molecules detected there (other than
CO and \treceCO).
Noticeably, strong lines of \ammonia\ and \hydroxil\ are observed, where 
hydroxil\ is the brightest of all detected lines in this source. Only a handful of water lines are detected 
on top of the fundamental levels of \owater\ and \pwater\ (see Table~\ref{tablines1}).
While the CO lines have roughly parabolic line profiles, the rest of the detected species display
a strong asymmetry in the form of a blue-shifted intensity excess. This feature is particularly clear 
in the \water\ and \ammonia\ profiles
(Fig.~\ref{fig_irc}, see also Menten et al. 2010), but only mildly observed in \hydroxil.
A similar trend was already reported in mm-wavelengths transitions of e.g. CS, \HtreceCN, or \vnsio\ by
Quintana-Lacaci et al. (2007). CC07 demonstrated that the main contribution of
this spectral component arises from a region in the western halo of the envelope (see their Fig.~8).
This isolated shell component was also reported in high-resolution mid-IR maps by Lagadec et al. (2011).
We propose that this spectral feature is due to an enhanced abundance of
certain species in these layers, as the model presented by CC07 predicts 
only moderate ($\sim$50\,K) temperatures at this radius.
\water\ and \ammonia\ are often believed to appear as a result of shock-induced chemistry,
suggesting that this gas layer could be composed of shocked material.

Section~\ref{model} presents a revised modelling of the whole set of
CO data and discusses the derived structure of the envelope around IRC+10420.

\subsection{AFGL\,2343}
\label{afgl}

\begin{figure}
\centering
\includegraphics[angle=0.0,width=\columnwidth]{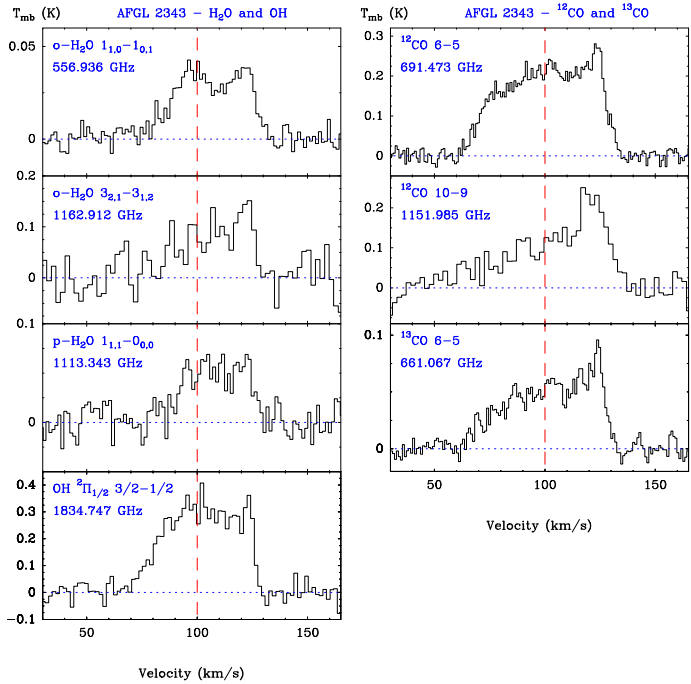}\\
\caption{Same as Fig.~\ref{fig_nml_h2o} for lines detected in AFGL\,2343.}
\label{fig_afgl}
\end{figure}

There is some controversy about the nature of AFGL\,2343, which has
been proposed by certain authors to be a post-AGB (low-mass) object
(see Ferguson \& Ueta 2010 and references therein), located at a
relatively short distance ($\sim$ 1 kpc, instead of $\sim$ 6 kpc if it
is a luminous hypergiant; note that no reliable parallax measurement of
this source exists). The majority of published papers favour the
massive-star hypothesis, in particular those reporting studies of the circumstellar shell
around AFGL\,2343 (Reddy \& Hrivnak 1999, Gledhill et al.\ 2001, CC07, Quintana-Lacaci et al.\ 2008), since the shell seems
very extended and similar to that of the well-studied IRC+10420. The 
fast and mostly isotropical expansion of its envelope 
also supports the YHG nature of AGL\,2343. Accordingly, we also assume that AFGL\,2343 is a YHG, of spectral type G5.
CC07 reported a variable and very high mass-loss rate, with values as high as
3$\times$10$^{-3}$\my\ in the past, but having shown a sharp drop to a still considerable $\sim$
4$\times$10$^{-5}$\my\ about 1200 years ago.

AFGL\,2343 exhibits the faintest set of submm lines of the sample
studied here. In particular, it is the only source where no emission is
detected in the \doceCO\ \jdsq\ and \treceCO\ \jdn\ transitions. This
is in line with the modelling results reported by CC07, where AFGL\,2343 appears as a detached shell, with
relatively low temperatures in general and low density in the central
layers. A similar picture is derived from high resolution mid-IR maps (Lagadec et al. 2011).
CC07 deduced a temperature of lower than about 20\,K for the densest shell and
between $\sim$300\,K and 30\,K for the much-less-dense inner
component. These low temperatures may however be underestimated given 
the detection of several transitions with energy levels near $\sim$300\,K (e.g. \doceCO\ \jdn, \owater\ \jkkul{3}{2}{1}{3}{1}{2} or \hydroxil).
The lack of dense enough warm layers in this source is corroborated
by the \doce\ \jdn/\jsc\ line and integrated line ratios ($\sim$0.58 and $\sim$0.5, respectively). 
These values are significantly smaller than the
corresponding line and integrated line ratios for IRC+10420, which reach $\sim$0.77 and $\sim$0.71 respectively,  
confirming that most of the circumstellar gas in AFGL\,2343 is cooler
than the dominant circumstellar shells in IRC+10420. 

The various line profiles have a red-shifted excess at  
v$_{\rm LSR}$ $\sim$\,125\,\kms. The contrast between this narrow
component and the emission around the star systemic velocity is
higher for higher excitation lines.  
The same behaviour was observed in the mm-wave lines of several molecules
(e.g. SiO, Quintana-Lacaci et al. 2007). In contrast, this excess tends to disappear in
lower excitation lines and is for example hardly seen in the \doceCO\ 
\juc\ and \jdu\ spectra (Bujarrabal et al. 2001). Interestingly, the \doce\ \jdn/\jsc\ line ratio at this velocity increases to $\sim$0.81, 
implying that there is a gas component in AFGL\,2343 with excitation 
conditions at least comparable to those present in IRC+10420.

The origin of this spectral component remains
unclear. In the interferometric maps of \doceCO\ \juc\ and \jdu\ by
CC07, this feature appears as a
compact clump located in the same direction as the central star and
moving almost at the extreme positive {\em LSR} velocity. It would
then correspond to a part of the envelope placed basically behind the
star. Quintana-Lacaci et al. (2008) studied this component in more detail 
using maps of $^{29}$SiO \jdu\ and HCN \juc\ emission.  They
concluded that this feature comes from a component at relatively small radius of
high density and temperature, and deduced it to be a
very clumpy, thin shell. These authors suggest that the high-excitation
condensations probably result from anisotropic shock propagation,
particularly by analogy with the case of IRC+10420, in which signs of
shocks running within the envelope were also found from SiO emission
maps. 

We speculate that the spectral feature found at 125 \kms\ could also
result from a shock between a hemisphere of the circumstellar shell and
nearby interstellar gas. It is remarkable that the {\em LSR} velocity
of AFGL\,2343 is positive and high in absolute value, $\sim$100 \kms,
so the object is receding from us at high velocity. On the other hand,
the low-$J$ CO observations (Bujarrabal et al.\ 2001) show that there is
interstellar emission at moderate {\em LSR} velocities, between 5\,\kms\ and
10\,\kms.
We cannot be sure whether this gas is really close to the YHG, but if
this were the case, the strong difference in the projected velocities
should produce a shock in the shell hemisphere placed beyond the star,
which would show an {\em LSR} velocity equal to approximately the
stellar velocity plus the circumstellar expansion velocity, i.e.\ about
125 \kms.

\section{Modelling of the CO emission in IRC+10420}
\label{model}

\subsection{The model}

CC07 developed a detailed model of the
structure and kinematics of the CSE around IRC+10420 based
on interferometric CO $J$=1--0 and $J$=2--1 maps, together with the sub-mm
observations of CO transitions up to $J$=6--5 performed by Teyssier et al.\ (2006).

\begin{figure}
\centering
\includegraphics[angle=0.0,width=8cm]{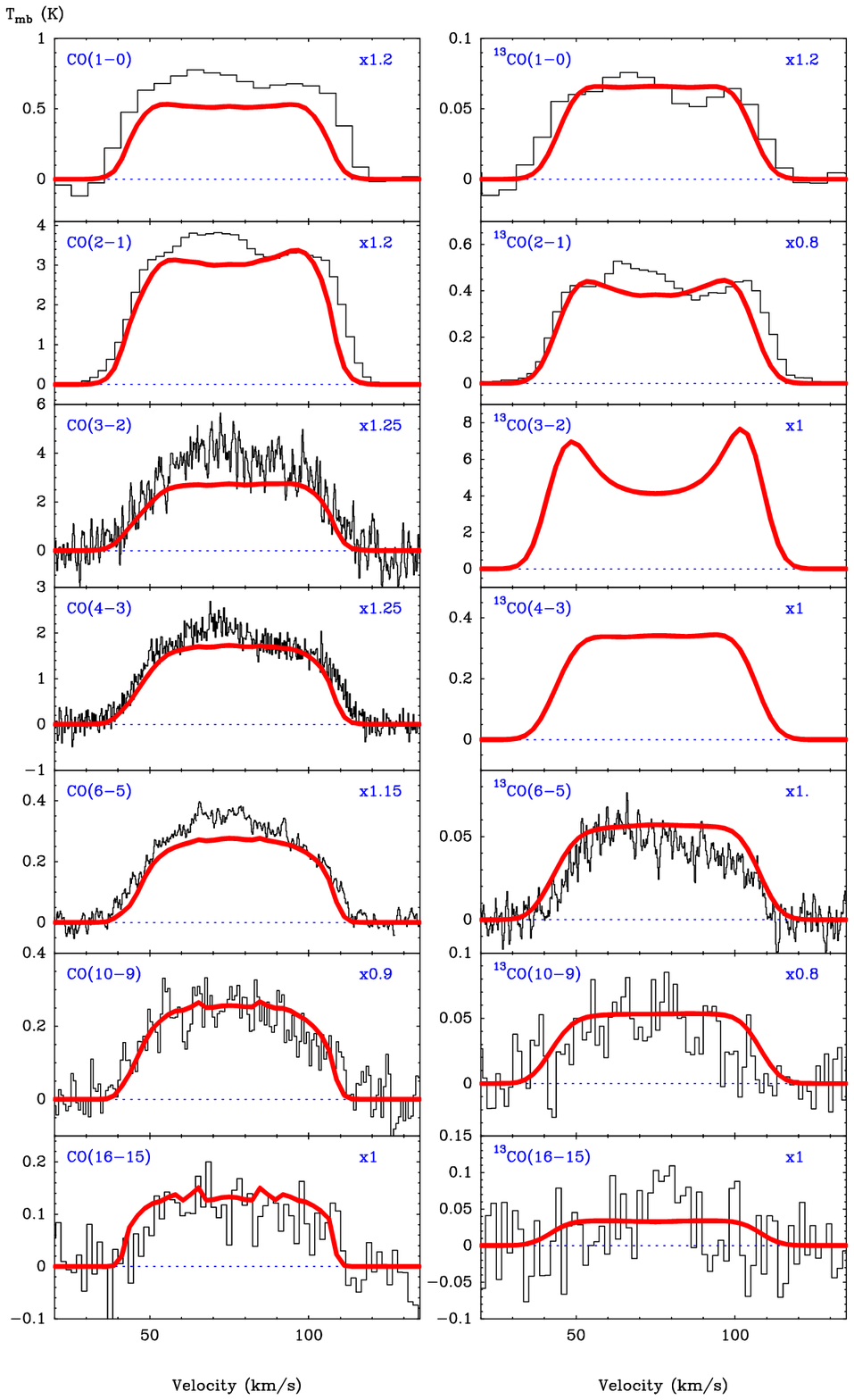}
\caption{Results of the model for IRC+10420 (red line) overlaid on
the line profiles presented in Fig.\ref{fig_irc}, as well as recent IRAM 30m data of the
\juc\ and \jdu\ transitions from Quintana-Lacaci et al. (in preparation) and the \doceCO\ \jtd\ and \jct\ transitions from the JCMT archive. Note that no data were found for the \jtd\ and \jct\ transitions of \treceCO. For each line, the model was adjusted by a free corrective factor of at most the absolute-calibration accuracy error, which was applied to the given transition. For the JCMT archive data, this calibration accuracy was unclear to us so we used a conservative upper limit of the 25\% error. For the IRAM 30m data, calibration errors of 20\% were assumed at both 3\,mm and 1\,mm. The correction factor is indicated in the upper right corner of each box.}
\label{fig_irc_model}
\end{figure}

\begin{figure}
\centering
\includegraphics[angle=0.0,width=\columnwidth]{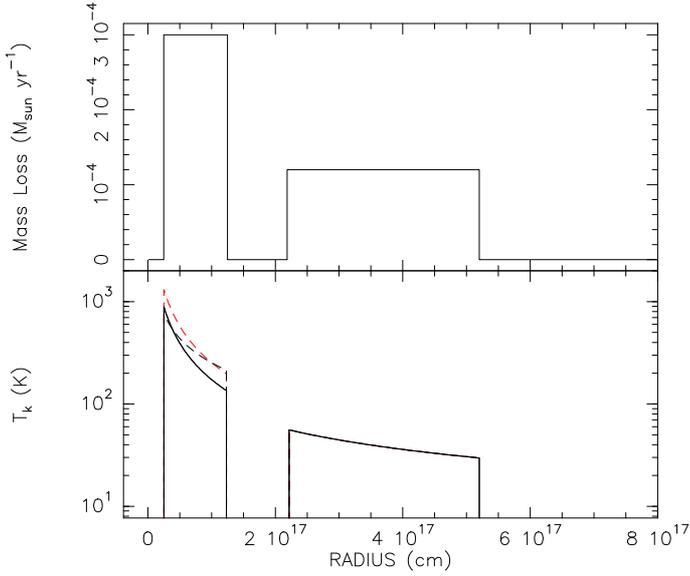}\\
\caption{{\it Upper panel}: Mass-loss pattern as a function of stellar radius derived for the model of IRC+10420 (Section~\ref{modelres}). {\it Lower panel}: Kinetic temperature profile as a function of stellar radius for the model of IRC+10420. The solid line corresponds to $T_{17}$ = 170\,K and $\alpha$ = $-$1.2, while the black dashed line uses $T_{17}$ = 230\,K and $\alpha$ = $-$0.8. The red dashed line illustrates the inner layer profile assumed in the model of CC07 ($T_{17}$ = 230\,K and $\alpha$ = $-$1.2).}
\label{fig_irc_model_profile}
\end{figure}

We performed numerical calculations of CO line excitation and
radiative transfer using a code very similar to that described in
CC07. Despite the evidence of departure from spherical
symmetry discussed in CC07 and Section~\ref{irc}, this model assumed isotropic mass
loss, and therefore used a one-dimensional code. 
The line excitation and the population of the rovibrational levels were
calculated using an large-velocity-gradient (LVG) approach, which is justified in view of the large
macroscopic velocities found in this object. The brightness
distribution for each line was calculated by solving the 'exact'
radiative transfer equations, so radiative interaction between distant
points and opacity effects were fully taken into account. The
brightness distribution was later convolved with the corresponding
telescope beams, which were assumed to be Gaussian. 

The original model by CC07 consisted of two detached shells,
each one corresponding to large mass loss episodes spaced at interval of 1200 years. It also assumed that the mass
loss ceased 200 years ago.
These respective mass ejections expand at high velocities (37\,\kms\ and 25\,\kms\ for the inner and outer shells,
respectively), as expected given the high luminosity of these objects.
Following CC07, the inner shell 
is assumed to be located between 2.5$\times$10$^{16}$
cm and 1.2$\times$10$^{17}$ cm and have a mass-loss rate of 3$\times$10$^{-4}$
\my. The outer shell is located between 2.2$\times$10$^{17}$ cm and 5.2$\times$10$^{17}$ cm 
and has a rate of 1.2$\times$10$^{-4}$ \my. The kinetic
temperatures were described by a power law $T_{\rm k}(r)=T_{17}
(r/10^{17} \rm{cm})^{\alpha}$, where $T_{17}$ is the temperature at
10$^{17}$ cm from the star. The gas temperatures derived by
CC07 were high, $T_{17}$ = 230\,K and $T_{17}$ =
100\,K for the inner and outer shells, respectively. Values of $\alpha$
$\sim$ $-$1 were found.  However, we note that low-$J$ transitions are not
good tracers of high-excitation regions, so that HIFI observations are
necessary to accurately constrain the temperature.

\subsection{Model results: Comparison with Herschel/HIFI data}
\label{modelres}

We compared the outcome of high-$J$ line profiles predicted
by this CSE model with our {\it Herschel}/HIFI data. Although that model was
derived from data of low-excitation lines, we find that it can fit all
the high-$J$ CO lines observed by HIFI by simply lowering the
kinetic temperature for the inner shell from $T_{17}$ = 230\,K to $T_{17}$ = 170\,K
in order to fit the \doceCO\ \jdsq\ line.  
This resulted in an average kinetic temperature of  $\sim$ 700\,K in the innermost part of the CSE.
We also note that the high-$J$ line intensities are fully
reproducible by restricting the model to an outer radius of 1.1$\times$10$^{17}$cm, i.e. roughly the
inner shell observed by CC07. This indicates that the outer shell around IRC+10420 is not probed by 
the {\it Herschel}/HIFI CO observations.

The result of the fitting of the HIFI $^{12}$CO and $^{13}$CO lines up
to $J$=16--15 from IRC+10420 is shown in Fig.\ref{fig_irc_model}.  
The corresponding mass-loss and kinetic temperature
profiles are displayed in Fig.\ref{fig_irc_model_profile}. 
We found that two different temperature distributions can reproduce the data
(Fig.\ref{fig_irc_model_profile}). In the first case, we adopted the same
(relatively steep) slope $\alpha$ = $-$1.2 as CC07, which led to
a global decrease in the temperature to $T_{17}$ = 170\,K. However,
we were also able to reproduce the observed lines we we considered a more gentle
increase in the temperature toward the centre, of $\alpha$ = $-$0.8 
and kept the original value of $T_{17}$ = 230\,K. 
In both models, the highest temperature of the circumstellar gas around
IRC+10420 remains \lsim\ 700 K, instead of the values \gsim\ 1000 K
originally assumed by CC07 (see Fig.\ref{fig_irc_model_profile}).

The assumed \doceCO\ abundance was 3$\times$10$^{-4}$
(typical of O-rich envelopes), and the \treceCO\ abundance was taken
from the study of Quintana-Lacaci et al. (2007). This translates into
a \doceCO/\treceCO\ abundance ratio of order 1/5, which is relatively
high compared to AGBs but very close to that reported in high-mass
post-red supergiants (see CC07, Quintana-Lacaci et al.\ 2007).  

\begin{figure}
\centering
\includegraphics[angle=270,width=\columnwidth]{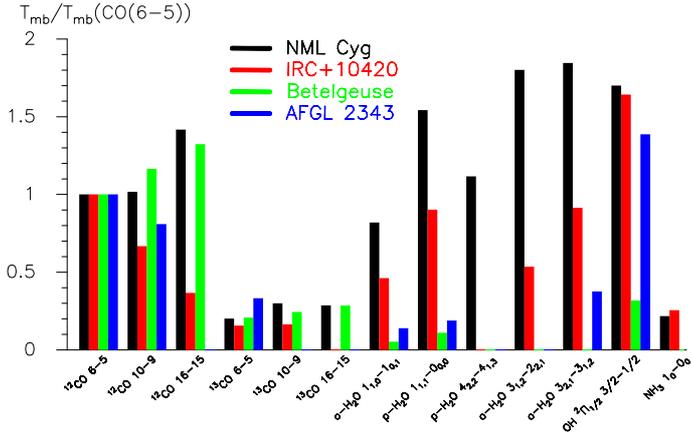}
\caption{Histograms of the peak brightness temperatures as tabulated in
Tables~\ref{tablines1} and~\ref{tablines2}, for lines detected in at
least 2 sources, normalised by the \doceCO\ \jsc\ intensity of the
respective sources. Non-detected lines are not represented in this graph.}
\label{fig_histo}
\end{figure}

\section{Conclusion}
\label{conclusion}

We have reported {\it Herschel}/HIFI observations of high spectral resolution for
FIR/submm molecular lines in two red supergiant stars (RSGs), NML Cyg
and Betelgeuse, and two yellow hypergiant stars (YHGs), IRC+10420 and
AFGL\,2343. 
Tables~\ref{tablines1} and~\ref{tablines2} summarize the observational
parameters derived from our data, which are displayed in Figs.~\ref{fig_nml_h2o} to~\ref{fig_afgl} -- the
full spectra are shown in the appendix.

As illustrated in Fig.~\ref{fig_histo},
the RSGs exhibit in general more intense high-excitation lines, indicating
that they harbour inner circumstellar layers with high temperatures
close to or even higher than 1000 K (Sections~\ref{nmlcyg} and~\ref{bet}). 
In contrast, YHGs do not contain this central warm material, and are instead 
surrounded by detached envelopes with low densities at small distances from the star
(Sections~\ref{irc} and~\ref{afgl}). As such, our results show that the mass-loss rates in 
YHGs are presently weak. 
We also observe that the spectral line profiles in RSGs tend to
become narrower with increasing excitation temperature. We show that
this phenomenon indicates that the high excitation lines in RSGs
originate from gas layers at small radii that have not yet reached the
circumstellar terminal velocity. 
This trend is not observed in YHGs, in line with the picture of a hollow 
envelope closer to the stellar photosphere. These YHGs are thus less likely than RSGs 
to contain gas still in the acceleration phase.

In both YHGs, we detected intense emission features that are conspicuous
in the profiles. That of IRC+10420 is relatively blue-shifted, at v$_{\rm LSR}$ $\sim$
65 \kms, and is more noticeable in certain molecules, such as H$_2$O and
NH$_3$, although it does not seem to require particular excitation conditions. 
It was detected and mapped in low-$J$ CO Lines by
CC07, who concluded that it comes from a
condensation in the outer envelope. We also found a peculiar emission excess
in AFGL\,2343 (Section~\ref{afgl}), at an extreme positive velocity of
v$_{\rm LSR}$ $\sim$ 125 \kms. There, the feature is
clearly associated with the excitation of the lines.
We speculate that this shell could result from a shock interaction
between the circumstellar envelope and nearby interestellar gas.

We stress the particularly strong emission of water lines in NML Cyg and IRC+10420,
which are also the only two sources featuring NH$_3$ emission.
On the other hand, OH is ubiquitously observed. It is unclear where this molecule
predominantly arises and how it forms in the envelopes. Its emission
does not seem to be correlated with that of \water, although OH formation in
evolved stars is thought to result from H$_2$O photodissociation; for
instance, the observed OH profiles are usually closer to those
of mid-excitation CO lines and the OH line intensity is high in
AFGL\,2343, which shows particularly weak water lines.

Finally, we have proposed a preliminary model to fit the
\doce\ and \trece\ lines detected in IRC+10420. 
The model is based on that developed by CC07 to explain their mm-wave interferometric mapping. 
We found that the original description of the shell around IRC+10420 by CC07
is able to reproduce all transitions by simply decreasing the temperature
of the inner layers by about 30\% (Fig.~\ref{fig_irc_model_profile}). In addition, the emission from the high-$J$ lines
is found to originate only from a detached, hot shell close to the star, formed by stellar
mass ejection at a high rate of $\sim$ 3$\times$10$^{-4}$ \my. Our calculation
indicates that no hotter components at small radii are needed to reproduce the
molecular emission of IRC+10420, confirming that the heavy mass-loss
ceased about 200 years ago.
More detailed modelling of the CO lines observed in IRC+10420 and
AFGL\,2343 will be presented in a forthcoming paper, in particular to
take into account the departure from spherical symmetry illustrated by
high-resolution maps at various wavelengths, and corroborated by the distinct
spectral features detected in the HIFI data.

\begin{acknowledgements}
The authors would like to thank the anonymous referee for constructive comments, allowing in particular 
this paper to become more concise.
HIFI has been designed and built by a consortium of institutes and university departments from across
Europe, Canada, and the United States under the leadership of SRON Netherlands Institute for Space
Research, Groningen, The Netherlands and with major contributions from Germany, France, and the US.
Consortium members are: Canada, CSA, U.Waterloo; France, CESR, LAB, LERMA, IRAM; Germany,
KOSMA, MPIfR, MPS; Ireland, NUI Maynooth; Italy, ASI, IFSI-INAF, Osservatorio Astrofisico di Arcetri-
INAF; Netherlands, SRON, TUD; Poland, CAMK, CBK; Spain, Observatorio Astron\'omico Nacional (IGN),
Centro de Astrobiolog\'ia (CSIC-INTA); Sweden, Chalmers University of Technology - MC2, RSS \& GARD,
Onsala Space Observatory, Swedish National Space Board, Stockholm University - Stockholm Observatory;
Switzerland, ETH Zurich, FHNW; USA, Caltech, JPL, NHSC.
      This work was supported by the German
      \emph{Deut\-sche For\-schungs\-ge\-mein\-schaft, DFG\/} project
      number Os~177/1--1. 
A portion of this research was performed at the Jet Propulsion Laboratory, California Institute of Technology, under contract with the National Aeronautics and Space Administration. RSz
and MSch ackowledge support from
grant N203 581040 of National Science Center.
\end{acknowledgements}

\newpage

\begin{appendix}
\section{Full HIFI spectra}
\label{appendix}

\begin{figure*}
\centering
\includegraphics[angle=0.0,width=15cm]{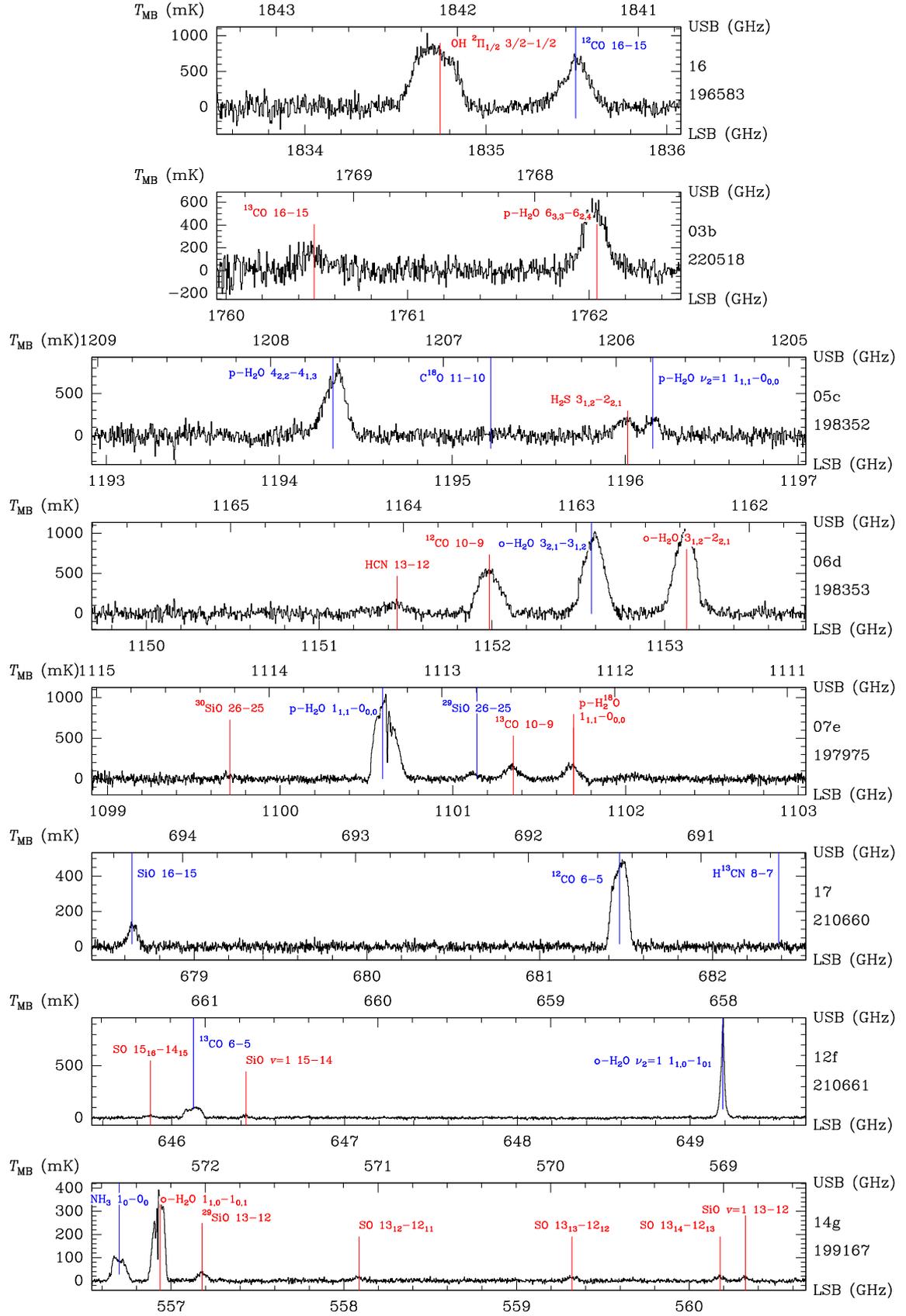}\\
\caption{Full Wide-Band Spectrometer spectra obtained towards NML\,Cyg. The setting number is indicated on the right side. Each spectrum is displayed on a double-side-band scale with both side-band frequency scales drawn. The spectral resolution is the native one (1.1\,MHz), and low-order polynomial baselines have been subtracted. Lines are indicated with the respective ticks at the expected (velocity-corrected) frequency -- lines from the USB have their ticks pointing upwards, while lines from the LSB have ticks pointing downwards.  The spectrum in setting 16 is made of only a fraction of the data taken at this frequency, for the purpose of standing wave mitigation (see text for details). The {\it obsid's} numbers are given below the setting number, in the form of 1342x, 'x' being the identifier indicated on the plot.}
\label{fig_nml_all}
\end{figure*}

\begin{figure*}
\centering
\includegraphics[angle=0.0,width=15cm]{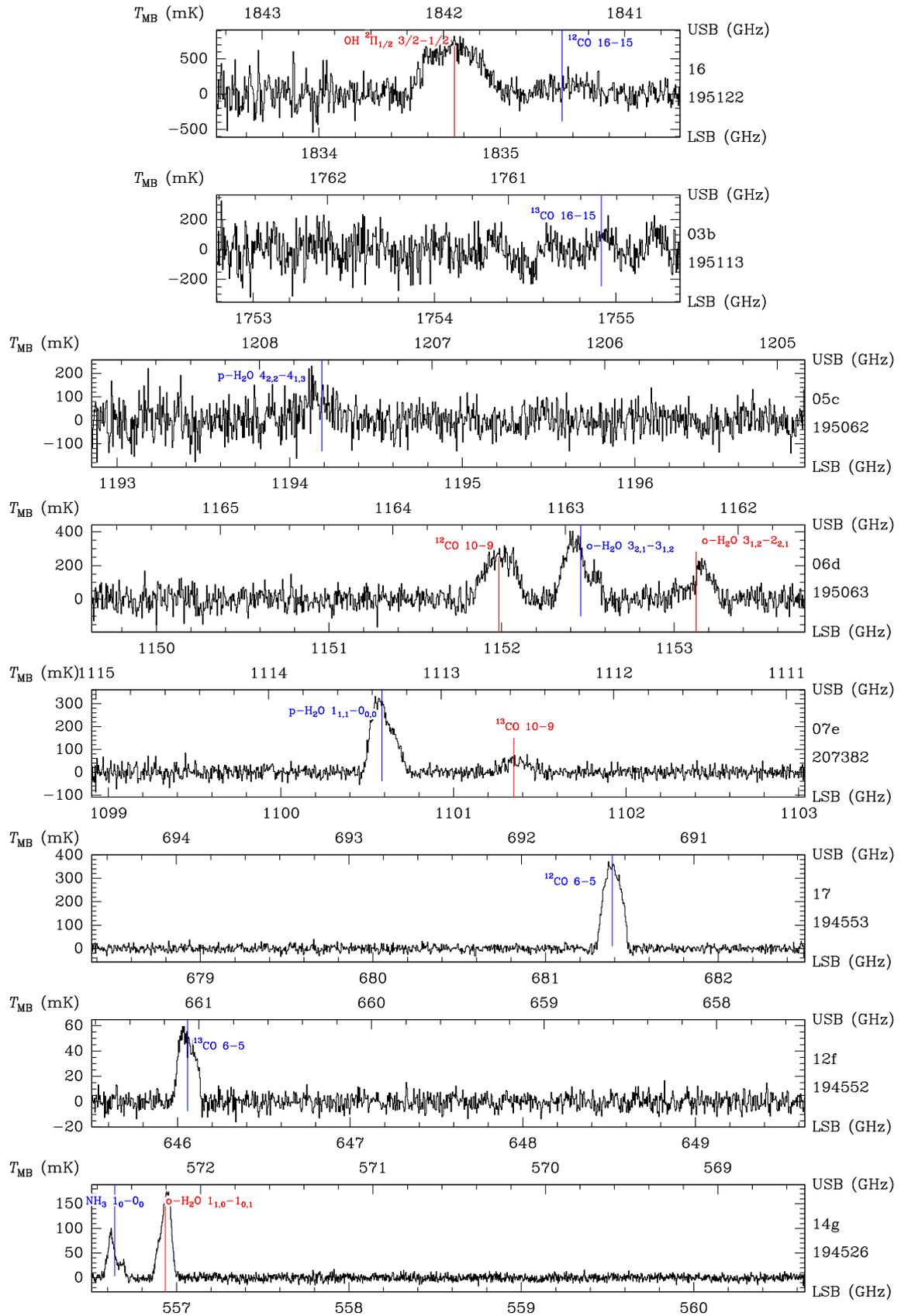}\\
\caption{Same as Fig.~\ref{fig_nml_all} for IRC+10420. The spectrum in setting 16x is made of only a fraction of the data taken at this frequency, for the purpose of standing wave mitigation (see text for details). The tick displayed for \treceCO\ \jdsq\ illustrates the expected position of this line, despite its non-detection.}
\label{fig_irc_all}
\end{figure*}

\begin{figure*}
\centering
\includegraphics[angle=0.0,width=15cm]{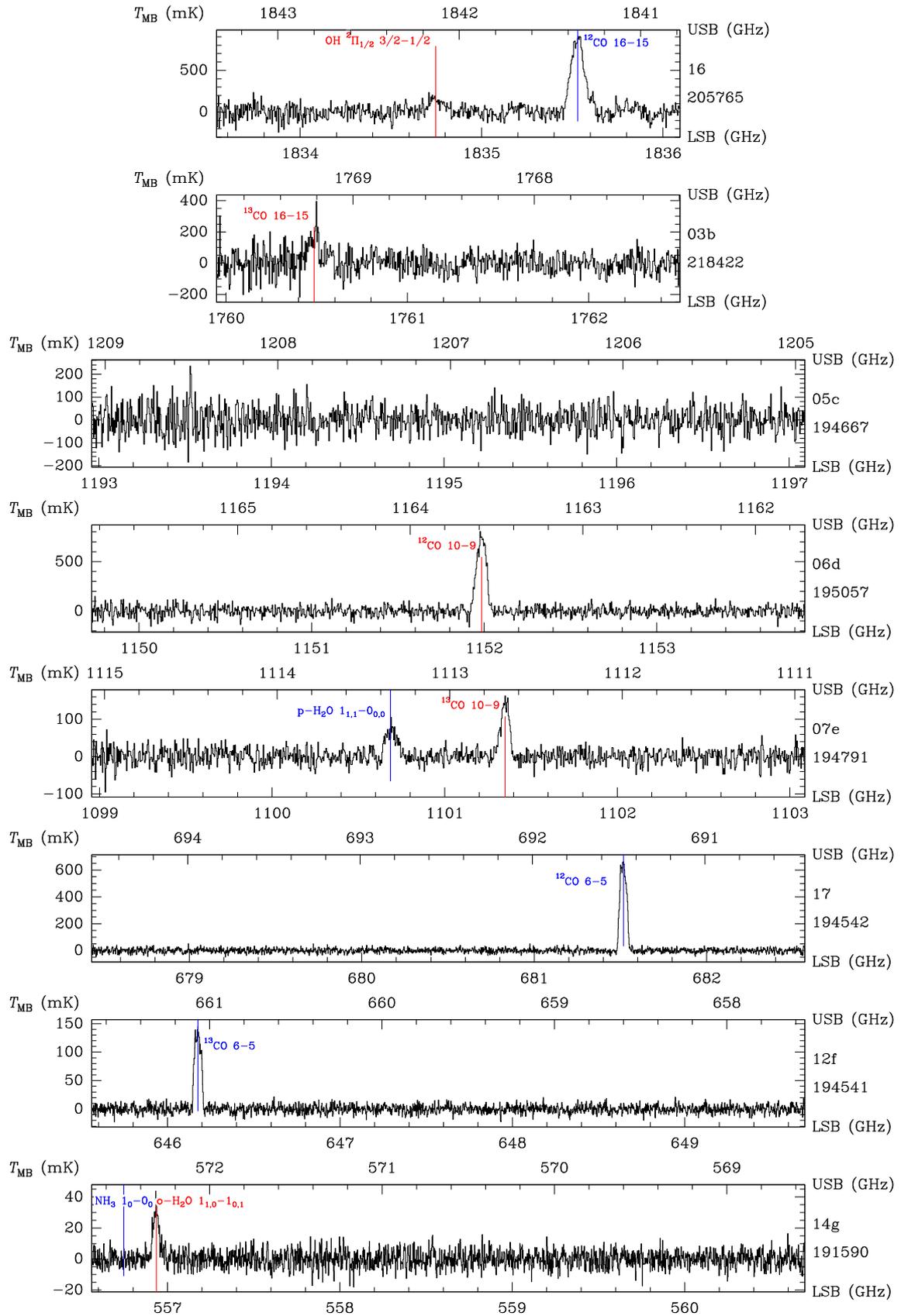}\\
\caption{Same as Fig.~\ref{fig_nml_all} for Betelgeuse. The tick displayed for \ammonia\ illustrates the expected position of this line, despite its non-detection.}
\label{fig_bet_all}
\end{figure*}

\begin{figure*}
\centering
\includegraphics[angle=0.0,width=15cm]{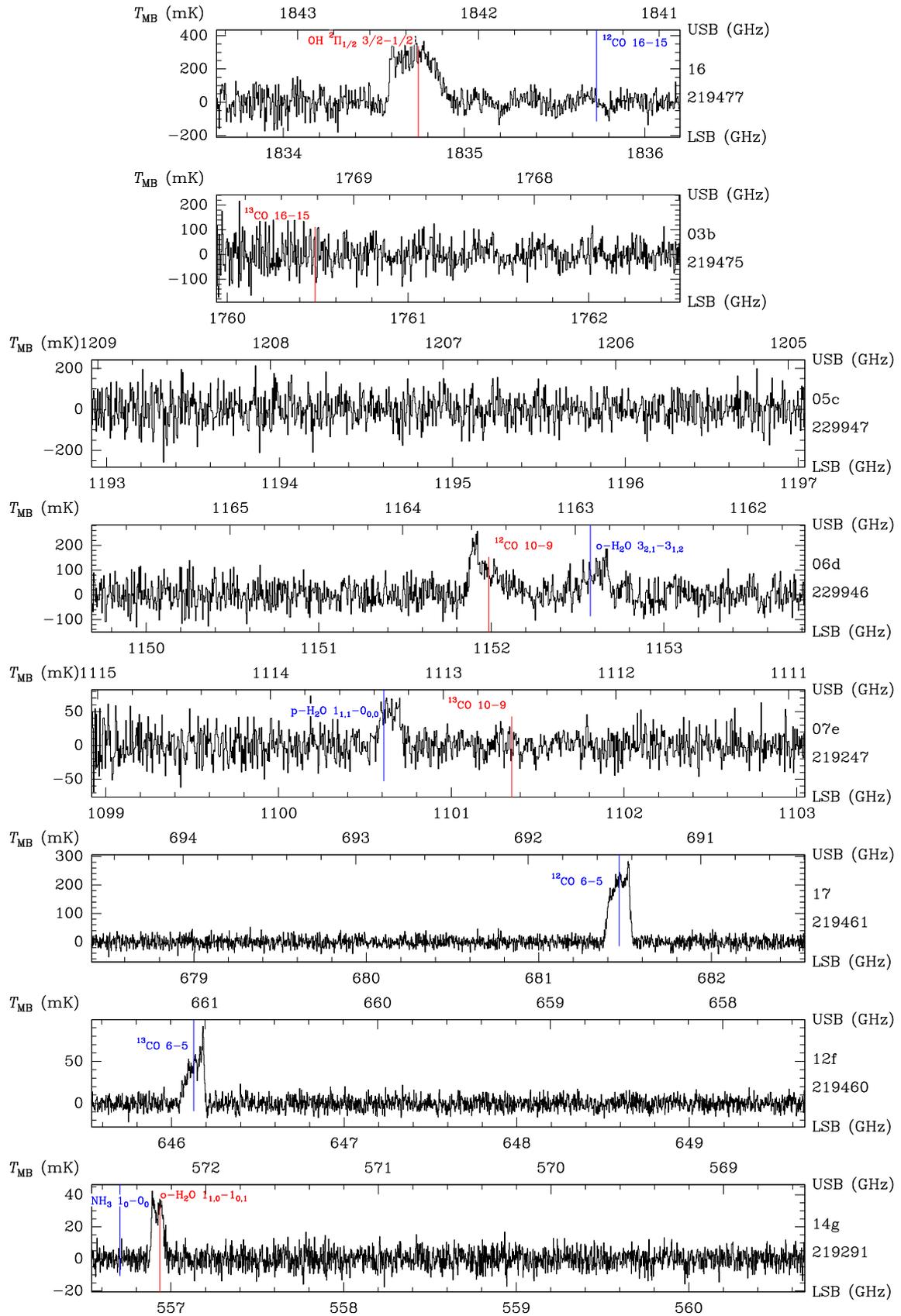}\\
\caption{Same as Fig.~\ref{fig_nml_all} for AFGL\,2343. The ticks displayed for \ammonia\ , \treceCO\ \jdn, \doceCO\ \jdsq\ and \treceCO\ \jdsq\ illustrate the expected position of these lines, despite their non-detection. }
\label{fig_afgl_all}
\end{figure*}

\end{appendix}

\end{document}